\documentclass[doublecol]{epl2}

\usepackage{amsfonts}
\usepackage{amsmath}
\usepackage{hyperref}
\usepackage{subcaption}
\usepackage{graphicx}
\usepackage{slashed}

\hypersetup{colorlinks=true,linkcolor=blue,citecolor=blue,filecolor=blue,urlcolor=blue,breaklinks=true}

\title{On the electromagnetic interaction and the anomalous term in the Duffin-Kemmer-Petiau theory}
\shorttitle{On the electromagnetic interaction and the anomalous term in the DKP theory} 

\author{Andr\'{e}s G. Jir\'{o}n\inst{1}\and Luis B. Castro\inst{2}\and Antonio S. de Castro\inst{3} \and Angel E. Obispo\inst{1,4}}
\shortauthor{Andrés G. Jirón\etal}

\institute{
  \inst{1} Facultad de Ingeniería, Universidad Tecnológica del Perú (UTP), Los Olivos, Lima, Perú\\
  \inst{2} Departamento de Física, Universidade Federal do Maranhão (UFMA), S\~{a}o Luís, MA, Brazil\\
  \inst{3} Departamento de Física, Universidade Estadual Paulista (UNESP), Guaratinguetá, SP, Brazil\\
  \inst{4} Departamento de Ciencias, Universidad Privada del Norte (UPN), Los Olivos, Lima, Perú}
\abstract{ The problem of vectorial mesons embedded in an electromagnetic field via Duffin-Kemmer-Petiau (DKP) formalism is reinvestigated. Considering the electromagnetic interaction as a minimal coupling, an incorrect value $(g=1)$ is identified for the gyromagnetic factor ($g$-factor). Furthermore, it is shown that is cumbersome to find analytical solutions due to the presence of the so-called anomalous term for the spin-1 sector of the DKP theory. Suspecting that the anomalous term results from an incomplete version of the DKP equation to describe the electromagnetic interaction, we consider the addition of a non-minimal coupling. This leads to the correct $g$-factor $(g=2)$, and as a consequence, the anomalous term becomes proportional to an external four current. As an application, the DKP equation with a static uniform magnetic field is considered, yielding the corresponding Landau levels.} 
\begin{document}

\maketitle

\section{INTRODUCTION}
\label{intro}
\noindent The gyromagnetic ratio ($g$-factor) of elementary charged particles coupled to an electromagnetic field is universally acknowledged to be $2$ for any spin, countering Belinfante's conjecture which proposed $g=1/s$ for particles with spin $s$ \cite{PR92:997:1953}. Initially confirmed by Weinberg \cite{DESER1970}, the value $g = 2$ was subsequently supported by Ferrara et al{.} \cite{PRD46:3529:1992} twenty-two years later, both employing non-minimal electromagnetic couplings. For instance, for spin-$1/2$ charged particles, Dirac theory with minimal electromagnetic coupling predicts $g=2$, which is agreement to the experimental values, except for a very small correction. Such correction, related to anomalous magnetic moment of the electron, is explained by the Pauli term (non-minimal electromagnetic coupling).

Nonetheless, the Proca theory with minimal coupling, as it pertains to spin-$1$ charged particles, predicts a value of $g=1$. Corben and Schwinger \cite{PR58:953:1940} raised concerns regarding this matter, which were addressed through modifications to the Proca Lagrangian, incorporating non-minimal coupling to obtain $g=2$. Various formalisms, including Sakata-Taketani \cite{JPA26:1397:1993}, Shay-Good \cite{PR179:1410:1969}, and electroweak part of the Lagrangian Standard Model \cite{PRD73:093009:2006} converge to a consensus of $g=2$ for spin-$1$ particles. From this, we can infer that a consistent theory for spin-$1$ charged particles must furnish $g=2$.


On the other hand, the Duffin-Kemmer-Petiau (DKP) theory \cite{petiau1936published,kemmer1938quantum,duffin1938characteristic,kemmer1939particle} has emerged as an alternative formalism to the established Klein-Gordon (KG) and Proca theories, offering a description of spin-$0$ and spin-$1$ particles with a richness of couplings inexpressible in competing theories \cite{PRD15:1518:1977,JPA12:665:1979}. The equivalence between these formalisms holds in the cases of free states and minimally coupled vector interactions \cite{UMEZAWA1956,PLA244:329:1998,PLA268:165:2000,PRA90:022101:2014}, ensured by a proper interpretation of the DKP spinor components (physical components). Regarding this matter, there are some subtle details that should be mentioned.

Nowakowski \cite{PLA244:329:1998} has shown that the solutions of the second order DKP equation are not always solutions of the first order DKP equation and the second order equation is just one member off a class of second order equations that can be obtained from DKP equation. This is closely related to the presence of an anomalous term in the DKP theory. Nowakowski has suggested a way of this problem may be circumvented. This way is based on the introduction of a higher (third) order wave equation of the DKP formalism. The lack of a back-transformation, which would allow us to obtain solutions of the first order DKP equation from solutions of the second order equation as it is in the Dirac theory does not give a possibility within the framework of the DKP theory to construct the path integral representation for the Green's function of a spin-$1$ particle in a background gauge field in a spirit of the approaches developed for a spin$1/2$ particle. This problem is discussed in detail in \cite{PRD92:105017:2015,JHEP07:94:2020} and ways to construct the path integral representation are proposed.
Considering the DKP equation interacting minimally with an electromagnetic field \cite{PLA268:165:2000}, an additional term appears, called the anomalous term, because it has no equivalent in the spin-$1/2$ Dirac theory. In Ref. \cite{PLA268:165:2000} was concluded that {\it ``when we select the physical components of DKP field $\psi$ the anomalous term is eliminated, so it has no physical meaning''}. Nevertheless, this statement is valid only for the spin-0 sector of the theory, assuming the same for the spin-1 sector would be a mistake given that the anomalous term persists even after selecting the physical components of the DKP spinor (second term in equation (21) in Ref.\cite{JPA43:495402:2010}). 

This misinterpretation of the conclusions in the seminal work \cite{PLA268:165:2000} was used for some authors to eliminate the anomalous term directly \cite{JPA43:495402:2010,JPA51:035201:2017} or indirectly (eliminating physical components) \cite{EPJC72:2217:2012,CJP91:1:2013,FP43:225:2013,ADHEP2014:185169:2014,EPJP130:236:2015,EPJP132:186:2017,CJP95:999:2017,EPL118:10002:2017,ChPB27:010301:2018,JMP60:013505:2019,PLA384:126706:2020,MPLA35:2050278:2020} which would lead to a loss of information of the physical system. Two questions remain: How can we deal with the anomalous term? Does the term anomalous have a physical meaning? These are questions that have not been rigorously addressed, and we aim to answer them in this work.

Here, we verify that employing minimal coupling for electromagnetic interactions for spin-1 sector of DKP theory yields a predicted value of $g=1$. However, as mentioned previously, a consistent theory for spin-1 charged particles is expected to predict a value of $g=2$. Therefore, it is reasonable to speculate that the presence of the anomalous term could be a result of an incorrect $g$-factor, i.e., a consequence of an incomplete version of the DKP equation to describe the interaction of spin-1 particles with the electromagnetic field. From Dirac theory, which associates the anomalous magnetic coupling with a second-order antisymmetric tensor interaction, we utilize this framework as a guide to describe non-minimal electromagnetic coupling within the DKP theory \cite{JPA12:665:1979}. We show that this modified DKP equation predicts a correct $g=2$, resulting in the anomalous term becoming proportional to an external four-current. As an application example, we reexamine the DKP equation (spin-1 sector) with a static uniform magnetic field and we performed the correct solution for this problem, which are in full agreement with those obtained via alternative formalisms \cite{JPA26:1397:1993,PR179:1410:1969,PRD73:093009:2006}.
Our work demonstrates that the modified DKP equation is a consistent theory for describing spin-1 charged particles, and establishes the equivalence with the generalized Proca equation \cite{PR58:953:1940} and the Standard Model \cite{PRD73:093009:2006}. This long-standing issue has now been resolved. Our findings create new opportunities by offering the modified DKP equation as an alternative model to the Standard Model. This offers the inherent advantage of describing a wide variety of interactions beyond the scope of the Standard Model. Consequently, we believe that the modified DKP equation holds significance in the quest for a more comprehensive understanding of spin-1 particles and their interactions.


\section{The Duffin-Kemmer-Petiau equation}
\label{section:DKP}
\noindent The first-order Duffin-Kemmer-Petiau (DKP) equation for a free boson of rest mass $M$ is given by \cite{kemmer1939particle} ($\hbar=c=1$)
\begin{equation}\label{ec-dkp}
(\beta^\mu \partial_\mu-M)\psi=0, 
\end{equation}
\noindent where the matrices $\beta^{\mu}$ satisfy the following algebra
\begin{equation}\label{algebradkp}
    \beta^\mu \beta^\nu \beta^\eta+\beta^\eta \beta^\nu \beta^\mu =g^{\mu \nu} \beta^\eta+g^{\eta \nu }\beta^\mu,
\end{equation}
\noindent and $g^{\mu \nu}$ is the Minkowski metric with the signature $(+,-,-,-)$. The algebra provides a set of $126$ independent matrices, which are part of three irreducible representations: $(i)$ a trivial representation (no physical meaning), $(ii)$ a five-dimensional representation (spin-$0$ sector), and $(iii)$ a ten-dimensional representation (spin-$1$ sector). To facilitate the identification of the physical components associated with the spin of each sector of the DKP theory, Umezawa \cite{UMEZAWA1956} proposed a set of projection operators that select the components of the spinor $\psi$ with well-defined Lorentz transformation properties: scalar, vector, and second-order tensors. Using these projection operators, the second-order Klein-Gordon (KG) and Proca equations are obtained when one selects the spin-$0$ and spin-$1$ sectors of the DKP theory; thus, the equivalence between theories is guaranteed in the free case \cite{UMEZAWA1956,PLA244:329:1998,PLA268:165:2000,PRA90:022101:2014}. Hereafter, we focus on the spin-$1$ sector of the DKP theory. 

To select the physical components of the DKP spinor for the spin-$1$ sector, we use the operators \cite{UMEZAWA1956}
\begin{equation}\label{r1}
    R^\mu \equiv (\beta^1)^2(\beta^2)^2(\beta^3)^2 (\beta^\mu \beta^0-g^{\mu 0}  ), 
\end{equation}
\noindent and 
\begin{equation}\label{r2}
R^{\mu\nu}=R^{\mu}\beta^{\nu}\,.
\end{equation}
\noindent The operators (\ref{r1}) and (\ref{r2}) satisfy the following properties

\begin{equation}
    R^{\mu\nu}=-R^{\nu\mu}\,,\label{r3}
\end{equation}
\begin{equation}
    R^{\mu \nu} \beta^\alpha = g^{\nu \alpha} R^\mu - g^{\mu \alpha} R^\nu, \label{r4}
\end{equation}
\begin{equation}
    R^\mu S^{\nu \alpha} = g^{\mu \nu} R^\alpha-g^{\mu \alpha}R^\nu , \label{r5}
\end{equation}
\begin{equation}
    R^{\mu \nu} S^{\alpha \rho} = g^{\mu \rho} R^{\nu \alpha}-g^{\mu \alpha} R^{\nu \rho}+g^{\nu \alpha} R^{\mu \rho}-g^{\nu \rho} R^{\mu \alpha} ,\label{r7}
\end{equation}
\noindent where
\begin{equation}
   S^{\mu \nu}=[\beta^\mu,\beta^\nu]. 
\end{equation}
\subsection{The DKP equation with minimal coupling}
\label{section:DKP:mc}
\noindent Considering the minimal vector interaction, the DKP equation can be rewritten as
\begin{equation}\label{dkp}
    \left(i\beta^\mu D_\mu-M\right)\psi=0\,,
\end{equation}
\noindent where $D_{\mu}=\partial_{\mu}+iqA_{\mu}$ denotes the covariant derivative. Thus, by applying the $R^{\nu}$ and $R^{\mu\nu}$ operators to the DKP equation (\ref{dkp}), the following equations are obtained
\begin{eqnarray}
&&R^\nu \psi=\frac{i}{M}D_\mu\left(R^{\nu \mu}\psi\right),\label{dkp_pro1}\\
&&R^{\nu \mu} \psi=\frac{i}{M}U^{\mu \nu},\label{dkp_pro2}
\end{eqnarray}
\noindent where $U^{\mu \nu}=D^\mu \left(R^\nu\psi \right)-D^\nu \left(R^\mu \psi\right)$. By substituting (\ref{dkp_pro2}) into (\ref{dkp_pro1}), we obtain
\begin{equation}\label{projf12-1}
( D_\mu D^\mu+M^2)(R^\nu \psi)+\frac{iq}{2}R^\nu S^{\alpha \sigma}F_{\alpha \sigma}\psi-D^\nu (D_{\mu}R^{\mu}\psi ) =0   . 
\end{equation}
\noindent with
\begin{equation}\label{subsi-1} 
 D_\mu R^\mu \psi=\frac{iq}{2M^2}F_{\mu\nu} U^{\mu\nu}, 
\end{equation}
\noindent where $F_{\mu \nu}$ is the electromagnetic tensor. Equation (\ref{subsi-1}) is called {\it subsidiary equation}, which relates the physical components of the spin-$1$ sector of the DKP theory. Thus, both the equation of motion (\ref{projf12-1}) and subsidiary equation (\ref{subsi-1}) are necessary to accurately describe the behavior of massive spin-$1$ particles interacting with an electromagnetic field via minimal coupling. It is noteworthy to mention that, upon performing algebraic manipulations, equation (\ref{projf12-1}) is reduced to the Proca equation minimally coupled to the electromagnetic field \cite{PLA268:165:2000,JPA43:495402:2010}. The last term in (\ref{projf12-1}) is called the anomalous term. Note that the anomalous term depends directly on the subsidiary equation, and this fact can indicate on the appearance of possible problems. As stated in Ref.\cite{JPA12:665:1979}, it is known that the minimal coupling has its own pathology in the DKP theory and we are forced to think that it is a consequence of the presence of the anomalous term. From (\ref{projf12-1}), one can see that even when selecting the physical components the anomalous term persists. In other words, the anomalous term contains physical information about the system and to eliminate it without any justification would lead to a loss of information. This misinterpretation of the conclusions in Ref.\cite{PLA268:165:2000} was used in Ref.\cite{JPA43:495402:2010}, where the authors studied the DKP equation with a uniform magnetic field for the spin-$1$ sector. Following the results of \cite{JPA43:495402:2010}, in Ref.\cite{JPA51:035201:2017} the authors studied the Aharonov-Bohm (AB) problem for vector bosons. On the other hand, a plausible possibility to eliminate the term anomalous in (\ref{projf12-1}) could be to require some restrictions on the form of the electromagnetic field. 

In the next section we will study the case of an external electromagnetic field and will explain in more detail the effects of the anomalous term on the dynamics of a spin-$1$ particle.
\section{The external electromagnetic field} Considering an external electromagnetic field and the representation of the electromagnetic tensor as $F^{0i}=-\mathcal{E}^{i}$ and $F^{ij}=-\epsilon^{ijk}B^{k}$, the subsidiary equation (\ref{subsi-1}) takes the following form 
\begin{align}
    &(E-qA_{0})\Phi-(\mathbf{P}-q \mathbf{A})\cdot \mathbf{\Psi} =\frac{iq}{M^2}\Big\{\bm{\mathcal{E}}\cdot
    \Big[(E-qA_{0})\mathbf{\Psi} \nonumber\\
    &-(\mathbf{P}-q \mathbf{A})\Phi \Big] -\mathbf{B}\cdot \left[(\mathbf{P}-q \mathbf{A})\times\mathbf{\Psi}\right]\Big\},\label{subs1}
\end{align}
\noindent where the physical components of the spin-$1$ sector have been written following the notation
\begin{align}
    \Phi&=R^{0}\psi,\\
    \mathbf{\Psi}^{T}&=(R^{1}\psi\,, R^{2}\psi\,, R^{3}\psi).
\end{align}
From (\ref{subs1}), one can see that it is possible to express component $\Phi$ in terms of the other three components $\mathbf{\Psi}$. This means that only three physical components are linearly independent and are related to the three degrees of freedom for a massive spin-$1$ particle. Therefore, we focus our attention  only on the linearly independent components of $\mathbf{\Psi}$.

Taking into account the same considerations and following some algebraic manipulation, equation (\ref{projf12-1}) is transformed into 
\begin{align}
    &\left[ (\mathbf{P}-q \mathbf{A})^2+M^2-(E-qA_{0})^2\right]\mathbf{\Psi}-q(\mathbf{S}\cdot \mathbf{B})\mathbf{\Psi} \nonumber\\
    &+iq\bm{\mathcal{E}}\Phi+\frac{iq}{M^2}(\mathbf{P}-q \mathbf{A})\Big\{\bm{\mathcal{E}}\cdot
    \left[(E-qA_{0})\mathbf{\Psi}-(\mathbf{P}-q \mathbf{A})\Phi \right]\nonumber \\
    &-\mathbf{B}\cdot\left[(\mathbf{P}-q \mathbf{A})\times\mathbf{\Psi}\right]\Big\}=\mathbf{0},\label{2do-rden1}
\end{align}
\noindent where the components of the spin operator $\mathbf{S}=(S^{1},S^{2},S^{3})$ are the $3\times 3$ spin-$1$ matrices
\begin{equation*}\label{spin3}
  S^1 = \left(\begin{array}{ccc}
     0 & 0& 0\\
 0& 0& -i\\
 0 & i&0\\ 
     \end{array}\right),\quad S^2 = \left(\begin{array}{ccc}
    0 & 0& i\\
 0& 0& 0\\
 -i & 0&0\\ 
 \end{array}\right), 
 \end{equation*}
 \begin{equation}\label{spin4}
 S^3 = \left(\begin{array}{ccc}
    0 & -i& 0\\
 i& 0& 0\\
 0 & 0&0\\ 
 \end{array}\right).
\end{equation}
\noindent The second term in (\ref{2do-rden1}) can be recognized as a Pauli-like term, essential for attributing physical meaning to the spin-dependent term. On the other hand, the anomalous term (last term) introduces the mixing of the physical components, significantly complicating the diagonalization of the equation system. Faced with this problem, some authors erroneously disregarded the anomalous term without justification \cite{JPA43:495402:2010,JPA51:035201:2017} or eliminated physical components to avoid its inclusion in the equation of motion \cite{EPJC72:2217:2012,CJP91:1:2013,FP43:225:2013,ADHEP2014:185169:2014,EPJP130:236:2015,EPJP132:186:2017, CJP95:999:2017,EPL118:10002:2017,ChPB27:010301:2018,JMP60:013505:2019,PLA384:126706:2020,MPLA35:2050278:2020}. To further explore the physical meaning of the anomalous term in this context, in the following subsection, we analyze the non-relativistic limit and $g$-factor of the DKP theory.

\subsection{The non-relativistic limit and gyromagnetic factor} Firstly, let us study the non-relativistic limit on the subsidiary equation. In this case, it is convenient to rewrite (\ref{subs1}) as
\begin{align}
    &(E-qA_{0})\Phi-(\mathbf{P}-q\mathbf{A})\cdot \mathbf{\Psi} =\frac{iqE}{M^2}\bm{\mathcal{E}}\cdot\mathbf{\Psi}-\frac{iq^{2}A_{0}}{M^2}\bm{\mathcal{E}}\cdot\mathbf{\Psi}\nonumber\\
    &-\frac{iq}{M^2} \mathbf{B}\cdot(\mathbf{P}\times\mathbf{\Psi})+\frac{iq^2}{M^2} \mathbf{B}\cdot(\mathbf{A}\times\mathbf{\Psi}). \label{subs_lnr}
\end{align}
\noindent In the non-relativistic limit (potential energy functions are smaller than $M$ and $E \sim M$), equation (\ref{subs_lnr}) becomes $\Phi\approx\frac{1}{M}\left(\mathbf{P}-q\mathbf{A}\right)\cdot\mathbf{\Psi}$, where the terms $\mathcal{O} \left(\frac{1}{M^2}\right)$ have been neglected. Thus, one can prove that at the low-speed regime ($|v|\ll 1$), one obtains $ \Phi \sim 0$. This last result reveals that the effects of component $\Phi$ are not perceived in the non-relativistic limit.

Let us now focus on equation (\ref{2do-rden1}), which in the non-relativistic limit transforms into
\begin{equation}\label{ecf1}
E_{NR}\mathbf{\Psi}=\left[\frac{1}{2M}(\mathbf{P}-q\mathbf{A})^2+qA_{0} -\frac{q}{2M} \mathbf{B}\cdot \mathbf{S}  \right]\mathbf{\Psi}\,,
\end{equation}
\noindent where $E_{NR}=E-M$ is the non-relativistic energy. From (\ref{ecf1}), one can see that the anomalous term naturally disappears in the non-relativistic limit and that $\mathbf{\Psi}$ obeys a Schrödinger-Pauli-like equation, except for the last term related to the $g$-factor, which in this case has $g=1$. This result is consistent with the well-known Belinfante conjecture \cite{PR92:997:1953}, which proposes a $g$-factor of $g=1/s$ for particles with arbitrary spin $s$.
However, as mentioned earlier, for spin-1 particles, there is currently a consensus favoring a $g$-factor equal to $g=2$, which has been verified in diverse contexts \cite{AJP74:1104:2006,marotta2021revisiting,EPL130:30003:2020,ponomarev2023basic}. Thus, one could think that the anomalous term could be a consequence of an incomplete version of the DKP equation to describe the interaction of spin-$1$ particles with the electromagnetic field (resulting in $g=1$). To complete the theory with $g=2$, it would be necessary to add an anomalous magnetic interaction (non-minimal electromagnetic coupling). In the next section, we will address this issue.

\section{Correction of the $g$-factor ($g=2$)}
\label{section:DKP:g2}

From Dirac theory, it is known that the anomalous magnetic coupling is associated to a second-order antisymmetric tensor interaction. This interaction in the DKP theory is proposed in \cite{JPA12:665:1979} as a linear combination of the only two second-order antisymmetric tensors and it is given by 
\begin{equation}\label{nominm}
    R=\frac{i q }{M}\left(S^{\mu\nu}-\frac{1}{4}\{\beta^\eta \beta_\eta,S^{\mu\nu}\}\right)F_{\mu\nu}\,.
\end{equation}
\noindent Due to the presence of derivatives of the vector potential in (\ref{nominm}), this interaction is considered non-minimal. It is worth noting that the anomalous magnetic interaction $R$ defined by (\ref{nominm}) does not violate causality, as demonstrated in \cite{JPA12:665:1979}.

Considering minimal and non-minimal electromagnetic couplings, the modified DKP equation becomes
\begin{equation}\label{proj1}
    \left[i\beta^\mu D_\mu-M+\frac{i q }{M}\left(S^{\mu\nu}-\frac{1}{4}\{\beta^\eta \beta_\eta,S^{\mu\nu}\}\right)F_{\mu\nu}\right]\psi=0\,.
\end{equation}
\noindent Following the same procedure as in the case with minimal coupling, we obtain (for more details, see Appendix \textbf{A})
\begin{align}
R^\nu \psi &=\frac{i}{M}D_\mu\left(R^{\nu \mu}\psi\right)-\frac{i q }{2M^2}R^\nu S^{\mu \sigma} F_{\mu \sigma}\psi,\label{proj3.1}\\
R^{\nu\mu}\psi &=\frac{i}{M}U^{\mu\nu}.\label{proj3.2}
\end{align}
\noindent Combining (\ref{proj3.1}) and (\ref{proj3.2}), we get
\begin{equation}\label{projf12}
( D_\mu D^\mu+M^2)(R^\nu \psi)+iqR^\nu S^{\mu\sigma}F_{\mu\sigma}\psi-D^\nu \left(D_\mu R^\mu \psi\right)=0  \,,    
\end{equation}
\noindent with
\begin{equation}\label{proj11}
        D_\mu R^\mu \psi=-\frac{iq}{M^2}J^{\mu}(R_{\mu}\psi)\,.
\end{equation}
\noindent where $J^{\mu}=\partial_{\nu} F^{\nu \mu}$ is the external four-current, which creates the external field $F^{\mu\nu}$. The equation obtained in (\ref{proj11}) is the \textit {subsidiary equation} for the electromagnetic case (considering a minimal and non-minimal coupling). The equations (\ref{projf12}) and (\ref{proj11}) describe completely a spin-$1$ particle in an electromagnetic field. It is worthwhile to mention that equation (\ref{projf12}) is in full agreement with the framework of a generalization of the Proca theory obtained by Corben and Schwinger (equation (19) in \cite{PR58:953:1940}) and also with \cite{PRD73:093009:2006}, which was obtained from the electroweak part of the Lagrangian of the Standard Model. The coefficient $1$ in front of the second term in (\ref{projf12}) corroborates that the $g$-factor takes the value $g=2$ and as a consequence of this the anomalous term becomes proportional to an external four-current. Therefore, if one considers for example regions outside those occupied by the external charges or a static uniform magnetic (electric) field, then the anomalous term will disappear. In the next section, we reexamined the case of an external electromagnetic field.  

\section{The external electromagnetic field reviewed}
\label{section:DKP:cmfr}

In this case, the {\it subsidiary equation} (\ref{proj11}) reads
\begin{equation}\label{subs_2}
        (E-qA_{0})\Phi-(\mathbf{P}-q\mathbf{A})\cdot\mathbf{\Psi}=\frac{q}{M^{2}}\left[\left(\mathbf{\nabla}\cdot\bm{\mathcal{E}}\right)\Phi -\left(\mathbf{\nabla}\times\mathbf{B}\right)\cdot\mathbf{\Psi}\right],
\end{equation}
\noindent and equation (\ref{projf12}) becomes
\begin{eqnarray}\label{ecf}
\left[ (\mathbf{P}-q\mathbf{A})^2+M^2-(E-qA_{0})^2\right]\mathbf{\Psi}-2q(\mathbf{S}\cdot\mathbf{B})\mathbf{\Psi}\nonumber\\
+\frac{q}{M^{2}}
\left( \mathbf{P}-q\mathbf{A} \right)\left[\left(\mathbf{\nabla}\cdot\bm{\mathcal{E}}\right)\Phi-\left( \mathbf{\nabla}\times\mathbf{B} \right)\cdot\mathbf{\Psi} \right]=\mathbf{0}\,.
\end{eqnarray}
\noindent The equation (\ref{ecf}) describes a spin-$1$ particle embedded in an external electromagnetic field via minimal and non-minimal couplings. Note that the second term in (\ref{ecf}) is a Pauli-like term with $g=2$. Therefore, one can conclude that the interaction $R$ given by (\ref{nominm}) really adjusts to the correct $g$-factor in the DKP theory. Furthermore, as a consequence of $g=2$, the anomalous term is proportional to an external four-current $J^{\mu}$. In this case, we have $J^{0}=\mathbf{\nabla}\cdot\bm{\mathcal{E}}$ and $\mathbf{J}=\mathbf{\nabla}\times\mathbf{B}$. Even after correcting the $g$-factor, the anomalous term makes it difficult to solve equation (\ref{ecf}); however, considering the case of a static uniform magnetic (electric) field, the anomalous term vanishes.

\subsection{The static uniform magnetic field}

Considering rectangular coordinates and choosing $A_{0}=0$ and $\mathbf{A}=(0,Bx,0)$, which furnishes a static uniform magnetic field in the $z$-direction, $J^{0}=\mathbf{\nabla}\cdot\bm{\mathcal{E}}=0$ and $\mathbf{J}=\mathbf{\nabla}\times\mathbf{B}=\mathbf{0}$, the equation (\ref{ecf}) reduces to
\begin{equation}\label{ecf2}
\left[ (\mathbf{P}-q\mathbf{A})^2+M^2-E^2-2qS^{3}B\right]\mathbf{\Psi}=\mathbf{0}\,. 
\end{equation}
\noindent Applying a similarity transformation in (\ref{ecf2}) and using 
\begin{equation}
 R^{i}\psi=\mathrm{e}^{ik_{y}y+ik_{z}z}R^{i}\varphi   ,
\end{equation}
 equation (\ref{ecf2}) becomes
\begin{equation}\label{amg83}
    \left[-\frac{1}{2M}\frac{d^2}{dx^2}+\frac{q^2B^2}{2M}\left(x-\frac{k_y}{qB}\right)^{2} \right]\chi^{s}= \varepsilon \, \chi^{s}, 
\end{equation}
\noindent where 
\begin{equation}
 \varepsilon = \frac{E^2-M^2+2qBs-k^2_z}{2M}   
\end{equation}
Here $\chi^{s}$ is eigenstate of the diagonalized spin operator $\bar{S}^{3}$ and satisfy $\bar{S}^{3}\chi^{s}=s\chi^{s}$, with $s=\pm1,0$ (for more details, see Appendix \textbf{B}). Note that equation (\ref{amg83}) takes on the form of the Schrödinger equation for the harmonic oscillator, with its eigenvalues corresponding to the well-known Landau levels \cite{bruckmann2018landau,atta2022finite}
\begin{equation}\label{mg93}
  E=\pm \left[M^2+k^2_z+qB\left(2n-2s+1\right) \right]^{1/2}. 
\end{equation}
\noindent In the above expression, the presence of factor $2qBs$ in the energy spectrum arises from the assignment of $g=2$. The expression (\ref{mg93}) is in good agreement with those of \cite{JPA26:1397:1993}, \cite{PR179:1410:1969}, and \cite{PRD73:093009:2006}, which were obtained using the Sakata-Taketani and Shay-Good formalisms, and from the electroweak part of the Lagrangian of the Standard Model, respectively.
\section{Conclusions}
\label{sec:fr}
We have reinvestigated the problem of interaction between vector mesons and an electromagnetic field within the framework of the Duffin-Kemmer-Petiau (DKP) formalism. Initially, considering the minimal electromagnetic coupling, we have identified an erroneous $g$-factor of $g=1$. Moreover, the presence of the anomalous term has significantly complicated the resolution of the DKP equation, with minimal electromagnetic interaction.
To address the $g$-factor discrepancy, we have introduced a non-minimal electromagnetic interaction. This additional term in DKP theory has been formulated as a linear combination of two second-order antisymmetric tensors. Consequently, considering the complete DKP equation with both minimal and non-minimal electromagnetic interactions has yielded the correct $g$-factor of $g=2$. As a result, the anomalous term has been redefined proportionally to the external four-current induced by the external field $F_{\mu\nu}$. Our findings remain robust, irrespective of the specific representation chosen for the DKP matrices.  Additionally, we have established an equivalence between the modified DKP equation, the generalized Proca equation \cite{PR58:953:1940}, and the Standard Model \cite{PRD73:093009:2006}. This underscores the consistency of the modified DKP equation in describing spin-1 charged particles within a theoretical framework.

Finally, we employed the modified DKP equation to reinvestigate the behavior of vector mesons in a static uniform magnetic field. We have mapped the modified DKP equation to a Schrödinger-like equation for the non-relativistic harmonic oscillator, obtaining an energy spectrum consistent with the established Landau levels. Our results are in full agreement with those obtained via other formalisms \cite{JPA26:1397:1993,PR179:1410:1969,PRD73:093009:2006}. In this sense, we would like to emphasize that our findings support the DKP theory as a alternative theory to the Standard Model for the description of a massive spin-$1$ field coupled to an electromagnetic field, with the bonus of having a wide variety of interactions that cannot be expressed in the framework of the Standard Model.

\section{Appendix A: Details for equations (\ref{proj3.1}), (\ref{proj3.2}) and (\ref{proj11})} \label{sec:appemdiceC}

By applying the $R^{\nu}$ and $R^{\nu\mu}$ operators to the DKP equation (\ref{proj1}), we obtain
\begin{equation}
\begin{split}
&(iR^{\nu\mu} D_\mu - MR^{\nu})\psi
\\&+ \frac{i q }{M}\left(R^{\nu}S^{\alpha\sigma}-\frac{1}{4}R^{\nu}\{\beta^\eta \beta_\eta,S^{\alpha\sigma}\}\right)F_{\alpha\sigma}\psi=0,
\label{1} 
\end{split}
\tag{A.1}
\end{equation}
\begin{equation}
\begin{split}
&(iR^{\nu\mu}\beta^\mu D_\mu - MR^{\nu\mu})\psi
\\&+ \frac{i q }{M}\left(R^{\nu\mu}S^{\alpha\sigma}-\frac{1}{4}R^{\nu\mu}\{\beta^\eta \beta_\eta,S^{\alpha\sigma}\}\right)F_{\alpha\sigma}\psi=0,
\label{2}
\end{split}
\tag{A.2}
\end{equation}

\noindent respectively. Using the operators properties, one can show that
\begin{equation}
    R^{\nu}\{\beta^\eta \beta_\eta,S^{\alpha\sigma}\} = 6R^{\nu}S^{\alpha\sigma}.
    \tag{A.3}
\end{equation}
\begin{equation}
    R^{\nu\mu}\{\beta^\eta \beta_\eta,S^{\alpha\sigma}\} = 4R^{\nu\mu}S^{\alpha\sigma}. \tag{A.4}
\end{equation}
\noindent Substituting these last results, the equations (\ref{1}) and (\ref{2}) becomes  
\begin{equation}
    \left(iR^{\nu\mu} D_\mu-MR^{\nu}-\frac{i q }{2M}R^{\nu}S^{\alpha\sigma}F_{\alpha\sigma}\right)\psi = 0\,,\label{5}
    \tag{A.5}
\end{equation}
\begin{equation}
    \left[i\left( D^{\mu}R^{\nu}-D^{\nu}R^{\mu} \right)-MR^{\nu\mu}\right]\psi = 0\,,\label{6}
    \tag{A.6}
\end{equation}
\noindent which can be expressed as in equations (\ref{proj3.1}) and (\ref{proj3.2}). \\

On the other hand, using the operator properties one can obtain the following relation
\begin{equation}\label{7}
R^{\nu}S^{\alpha\sigma}F_{\alpha\sigma}\psi=2F^{\nu\sigma}R_{\alpha}\psi\,. 
\tag{A.7}
\end{equation}
\noindent Using this last result, the equation (\ref{proj3.1}) can be rewritten as
\begin{equation}\label{8}
R^{\nu}\psi=\frac{1}{M^{2}}D_{\mu}U^{\nu \mu}-\frac{i q }{M^2}F^{\nu \sigma}R_{\sigma}\psi\,.
\tag{A.8}
\end{equation}
\noindent By applying $D_{\nu}$ to (\ref{8}), we get
\begin{equation}\label{9}
D_{\nu}R^{\nu}\psi=\frac{1}{M^{2}}D_{\nu}D_{\mu}U^{\nu \mu}-\frac{i q }{M^2}D_{\nu}\left(F^{\nu \sigma}R_{\sigma}\psi\right)\,.
\tag{A.9}
\end{equation}
\noindent Using
\begin{equation}
    D_{\nu}D_{\mu}U^{\nu \mu} = iqF^{\mu\nu}D_{\mu}R_{\nu}\psi\,,
    \tag{A.10}
\end{equation}
\begin{equation}
    D_{\nu}\left(F^{\nu \sigma}R_{\sigma}\psi\right) = \partial_{\nu}F^{\nu\sigma}\left(R_{\sigma}\psi\right) + F^{\nu\sigma}D_{\nu}R_{\sigma}\psi\,,
\tag{A.11}
\end{equation}
\noindent the equation (\ref{9}) becomes
\begin{equation}
D_{\nu}R^{\nu}\psi=-\frac{i q }{M^{2}}\partial_{\nu}F^{\nu\sigma}\left(R_{\sigma}\psi\right)\,,
\tag{A.12}
\end{equation}
\noindent which is the equation (\ref{proj11}).

\section{Appendix B: Diagonalization} \label{sec:appendice}

Multiplying from the left by the inverse matrix $D^{-1}$ and using 
\begin{equation}
 R^{i}\psi=\mathrm{e}^{ik_{y}y+ik_{z}z}R^{i}\varphi ,   
\end{equation}
the equation (\ref{ecf2}) becomes
\begin{equation}\label{ap1}
    \left[-\frac{d^2}{dx^2}+q^2B^2\left(x-\frac{k_y}{qB}\right)^{2}+M^{2}-E^{2}+k_{z}^{2}-2qB\bar{S^{3}} \right]\bm{\chi}=0,  \tag{B.1} 
\end{equation}
\noindent where
\begin{equation}\label{sz_diag}
\bar{S^{3}}=D^{-1}S^{3}D=
\begin{pmatrix}
    1 & 0& 0\\
 0& -1& 0\\
 0 & 0 &0\\ 
\end{pmatrix}\,,
\tag{B.2} 
\end{equation}
\noindent with
\begin{equation}
D=\begin{pmatrix}
    1 & 1 & 0\\
 i & -i & 0\\
 0 & 0 & 1\\ 
\end{pmatrix}\,, \quad
D^{-1}=\begin{pmatrix}
    1/2 & -i/2 & 0\\
 1/2 & i/2 & 0\\
 0 & 0 & 1\\ 
\end{pmatrix}\,. \tag{B.3}
\end{equation}
\noindent Here, the new spinor $\bm{\chi}$ can be defined as
\begin{equation}\label{newspinor}
\bm{\chi}=\begin{pmatrix}
    \chi^{+1}\\[0.2cm]
    \chi^{-1}\\[0.2cm]
    \chi^{0}\\ 
\end{pmatrix}=
\begin{pmatrix}
   \frac{R^{1}\varphi-iR^{2}\varphi}{2} \\[0.2cm]
 \frac{R^{1}\varphi+iR^{2}\varphi}{2} \\[0.2cm]
  R^{3}\phi \\ 
\end{pmatrix}\,, \tag{B.4} 
\end{equation}  
\noindent and each component is an eigenstate of $\bar{S^{3}}$ satisfying $\bar{S}^{3}\chi^{s}=s\chi^{s}$, with $s=\pm1,0$. In this way, we can rewrite the equation (\ref{ap1}) in a compact form as in (\ref{amg83}).

\acknowledgments
This work was supported in part by means of funds provided by CNPq, Brazil, Grants No. 09126/2019-3 and 311925/2020-0, FAPEMA and CAPES - Finance code 001. Angel E. Obispo acknowledges the financial support from the Universidad Tecnológica del Perú (UTP).

\bibliographystyle{spphys} 

\end{document}